\documentclass[sigconf]{acmart}

\usepackage{booktabs} 
\usepackage{enumitem}
\usepackage{subcaption}
\usepackage[utf8]{inputenc}
\setcopyright{rightsretained}
\usepackage{xcolor}

\acmDOI{}

\acmISBN{}

\acmConference[arXiv]{}{Kevin Jasberg and Sergej Sizov}{2017} 
\acmYear{2017}
\copyrightyear{2017}

\acmPrice{}

\begin{document}
\title{Re-Evaluating the Netflix Prize - \\ Human Uncertainty and its Impact on Reliability}
\titlenote{Full dataset and evaluation routines available at https://jasbergk.wixsite.com/research}

\author{Kevin Jasberg}
\affiliation{%
  \institution{Web Science Group\\Heinrich-Heine-University Duesseldorf}
  \city{Duesseldorf} 
  \state{Germany} 
  \postcode{45225}
}
\email{kevin.jasberg@uni-duesseldorf.de}

\author{Sergej Sizov}
\affiliation{%
  \institution{Web Science Group\\Heinrich-Heine-University Duesseldorf}
  \city{Duesseldorf} 
  \state{Germany} 
  \postcode{45225}
}
\email{sizov@hhu.de}

\renewcommand{\shortauthors}{}

\begin{abstract}
In this paper, we examine the statistical soundness of comparative assessments within the field of recommender systems in terms of reliability and human uncertainty. From a controlled experiment, we get the insight that users provide different ratings on same items when repeatedly asked. This volatility of user ratings justifies the assumption of using probability densities instead of single rating scores. As a consequence, the well-known accuracy metrics (e.g. MAE, MSE, RMSE) yield a density themselves that emerges from convolution of all rating densities. When two different systems produce different RMSE distributions with significant intersection, then there exists a probability of error for each possible ranking. As an application, we examine possible ranking errors of the Netflix Prize. We are able to show that all top rankings are more or less subject to high probabilities of error and that some rankings may be deemed to be caused by mere chance rather than system quality.
\end{abstract}

%
%


\maketitle

\section{Introduction}
Recommender systems play a central role nowadays and their sound evaluation is crucial.
For this purpose, a variety of quality metrics have been developed \cite{Herlocker}, such as the RMSE which has been used in one of the largest recommender competitions, the Netflix Prize. In this contribution we draw attention to possible inaccuracies within recommender assessment caused by uncertain user feedback, exemplary in the evaluation of the Netflix Prize.

In a systematic experiment, we required participants to (re-)rate theatrical trailers several times. Our results reveal that users are not able reproduce their own decisions, i.e. given ratings fluctuate around a central tendency. This result is consistent with other studies \cite{Hill} and theoretical models of the human mind \cite{Friston}. Based on our experiment and in accordance to the Netflix Prize, one may compute the RMSE for different recommender systems for each of the rating trials. 
\begin{figure}[t]
\centering
\includegraphics[width=.88\linewidth]{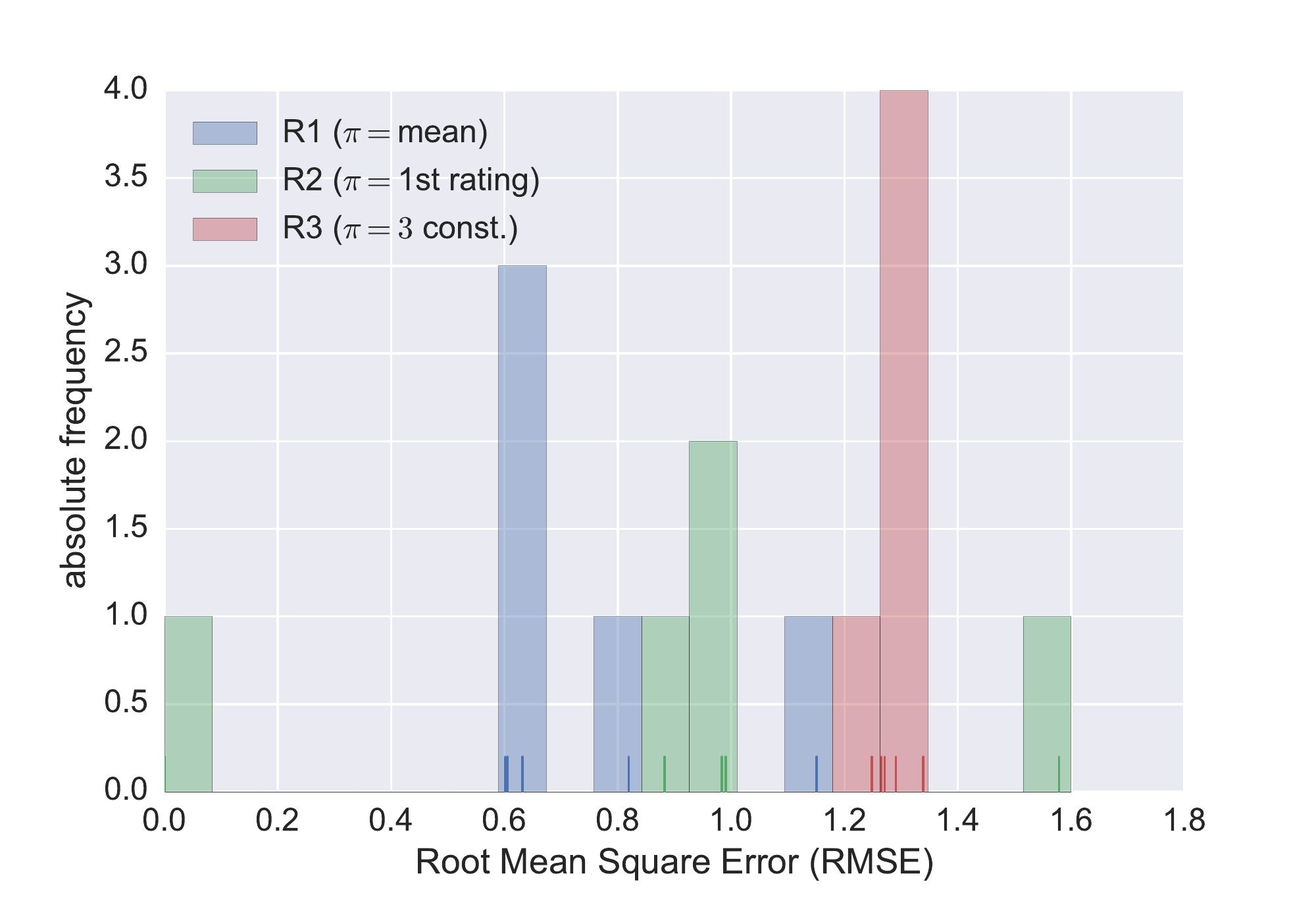}
\caption{Histogram of RMSE outcomes for three recommender systems in five repeated rating trials.\vspace{-6ex}}
\label{fig:DiscreteRMSE}
\end{figure}
Figure \ref{fig:DiscreteRMSE} shows a histogram of these different RMSE outcomes for three sample recommender systems (defined by their predictors $\pi$). It is apparent that the RMSE itself yields a particular degree of uncertainty, due to uncertain user feedback. When ranking these recommender systems, Figure \ref{fig:DiscreteRMSE} allows for a variety of possible orders that emerge with different frequencies.
The problem is most obvious for recommender R2 (green) as it could be both, the best or the worst recommender, although it operates on the same users rating the same items. Thus, the question for a comparison changes, namely from ``Is R1 better than R2?'' to ``How likely is it that R1 is better than R2?''. Vice versa, no matter what ranking we finally opt for, there is always a certain chance of error for this decision. 
The impact of uncertain user feedback and possible ranking errors is in the main focus of this paper and will be exemplified using the Netflix Prize. 

The central research question is thus: How reliable is the Netflix Prize (as an example for evaluations in general) when considering human uncertainty?

\section{Related Work}

The observation of uncertain user feedback in product evaluations was been made before in \cite{Hill}. 
The concept of this study has been combined with modern methods of experimental psychology \citep{psycho} to conduct out our own study.
Latest neuroscience research considers action-coordinating cognitions to be based on perceptions in the form of distributions which are constantly updated by a complicated generative process within the human cortex \cite{Friston}. Decision making thus yields a specific volatility, which we denote human uncertainty in our context. This uncertainty can be explained by the irregular release of neuromodulators like dopamine and acetylcholine \cite{FristonNature}. These findings support our idea of modelling user feedback as individual distributions.
The handling of uncertainty has a long tradition in the field of physics and metrology \cite{Ku,GUM,GUMsupp1}.
In particular, \cite{GUM} describes the propagation of uncertain quantities when new ones are calculated therefrom.
This model of uncertainty is used to calculate the distributions of the RMSE.
With this collection of methods, we are able to determine the human uncertainty experimentally, to investigate their propagation in the RMSE, and to uncover possible ranking errors in the Netflix Prize.

\section{Case Study} 
\begin{figure}[t]
\centering
\includegraphics[width=.92\linewidth]{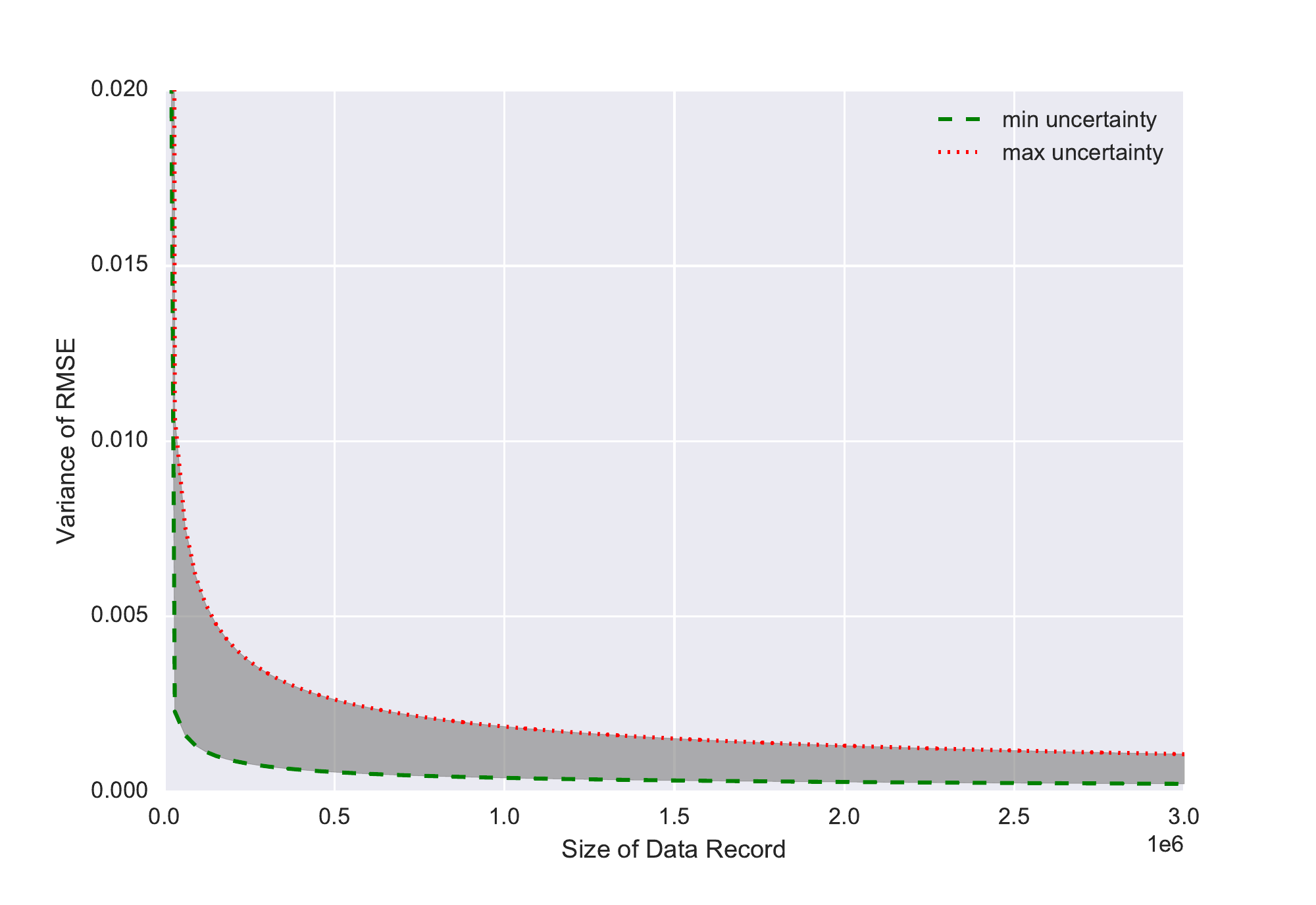}
\caption{Behaviour of the RMSE's variance as a function of data size and human uncertainty. \vspace*{-2ex}}
\label{fig:RMSEvariance}
\end{figure}

Let $X_\nu  \sim\mathcal{N}(\mu_\nu,\sigma_\nu^2)$  be a family of $n$ random variables (representing user ratings) which are assumed to be normally distributed in accordance to \cite{delia}. The RMSE thus becomes a random variable itself. The distribution emerges as a convolution of $n$ density functions with respect to the mathematical model
\begin{small}
\begin{equation} \label{eq:RMSE}
\operatorname{RMSE} = \sqrt{\frac{1}{n} \sum_{\nu} (X_\nu  - \pi_\nu )^2}.
\end{equation}
\end{small}
\hspace{-1.5ex} Using the Gaussian Error Propagation \cite{Ku} and the Central Limit Theorem, the $\operatorname{RMSE}\sim \mathcal{N}(\mu,\sigma^2)$  yields a normal distribution with
\begin{equation}
\mu\approx\sqrt{\frac{1}{n} \textstyle{\sum_{\nu}} \sigma_\nu^2+\Delta_\nu^2}
\quad\text{and}\quad
\sigma^2\approx\frac{\textstyle{\sum_{\nu}}\sigma_\nu^4+2\sigma_\nu^2\Delta_\nu^2}{2n\cdot \textstyle{\sum_{\nu}}\sigma_\nu^2+\Delta_\nu^2} .
\label{eq:RMSEparameters}
\end{equation}
with the substitution $\Delta_\nu=\mu_\nu-\pi_\nu$.
Let now $Z_1\sim \mathcal{N}(\mu_1,\sigma_1^2)$ and $Z_2\sim \mathcal{N}(\mu_2,\sigma_2^2)$ be two RMSE random quantities that correspond to different recommender systems. Assuming $\mu_1< \mu_2$, we would consider system 1 to be better than system 2. However, this decision may be subject to an error which occurs with a probability of
\begin{small}
\begin{equation}
P(Z_1\geq Z_2) 
= \Phi\,\left( (\sigma_1^2+\sigma_2^2)^{-1/2}(\mu_1-\mu_2)  \right). \label{eq:error}
\end{equation}
\end{small}
\hspace{-1.5ex} where $\Phi$ is the standard-normal cumulative distribution function.
With this framework we are able to elaborate the reliability of the Netflix Prize.
At this point, it appears to be challenging that Netflix did not collect any information about human uncertainty.
However, for the size of Netflix's test record ($n=2.8\cdot 10^6$), this is not a problem at all since the RMSE's variance scales with $1/2n$. This is illustrated in Figure \ref{fig:RMSEvariance}. It is apparent that the true extent of human uncertainty no longer influences the variance significantly when one has to deal with big data.
In fact, we estimated the uncertainty for the Netflix Prize in three different ways:
\begin{small}
\begin{description}
\item[Approach A)] ML-fitting of human uncertainty based on our experiment provided a density from which random draws were made to be associated to each rating of the Netflix record.
\item[Approach B)] Human uncertainty was randomly sampled from different distributions (e.g. uniform, triangular, beta) and associated to each rating of the Netflix record.
\item[Approach C)] Having a 5-star scale, human uncertainty yields certain limitations. Association of minimum and maximum uncertainty to each Netflix rating produces an interval in which the RMSE's variance is located.
\end{description}
\end{small}
With \ref{eq:RMSEparameters} we can then transform each RMSE $score$ in the Netflix leaderboard into a random quantity $Z\sim\mathcal{N}(score, \sigma_{score}^2)$. In doing so, methods A and B always provide the same value $\sigma_{score}^2= 0.0006$. For method C, there are intervals whose mean exactly corresponds to the result of method A and B. 
This empirically shows that the extent of human uncertainty for each individual rating no longer contributes to the variance of the RMSE since only the size of the data record is decisive here. With \ref{eq:error} we can then estimate the error probabilities that correspond to each pair-wise ranking.

The results are listed in the Table below.
$R_i$ represents the recommender system with leaderboard placing $i$.
The entry $p_{ij}$ is the error probability of the ranking $R_i < R_j$.
For example, the error probability of placing 3 being better than placing 4 is nearly 25\%, i.e. these systems would swap placings on the leaderboard in one of four repeated evaluations. Especially for the last placings there is a disillusioning message: Placings 9, 10, 11 and 12 hold nearly 50\% probability of error. Thus, the entry into the top 10 of the Netflix Price might be based on mere chance rather than system quality.\\[.25ex]
\hspace*{-2ex}\begin{footnotesize}
\begin{tabular}{lrrrrrrrrrrr}
&  $R_{1/2}$ & $R_3$ & $R_4$ &  $R_5$ &  $R_6$ &  $R_7$ &  $R_8$ &  $R_9$ &  $R_{10}$ &  $R_{11}$ &  $R_{12}$ \\
$R_{1/2}$  &  .50 
&      \textcolor{teal}{.04} 
&      \textcolor{teal}{.01} 
&      \textcolor{lightgray}{.00} 
&      \textcolor{lightgray}{.00} 
&      \textcolor{lightgray}{.00} 
&      \textcolor{lightgray}{.00} 
&      \textcolor{lightgray}{.00} 
&      \textcolor{lightgray}{.00} 
&       \textcolor{lightgray}{.00} 
&       \textcolor{lightgray}{.00} 
\\
$R_3$   &    &      .50 
&      \textcolor{red}{\textbf{.24}} 
&      \textcolor{orange}{.14} 
&      \textcolor{teal}{.08} 
&      \textcolor{teal}{.01} 
&      \textcolor{lightgray}{.00} 
&      \textcolor{lightgray}{.00} 
&      \textcolor{lightgray}{.00} 
&       \textcolor{lightgray}{.00} 
&       \textcolor{lightgray}{.00} 
\\
$R_4$   &    &    &      .50
&      \textcolor{red}{\textbf{.36}} 
&      \textcolor{red}{\textbf{.24}} 
&     \textcolor{teal}{.06} 
&      \textcolor{lightgray}{.00} 
&      \textcolor{lightgray}{.00} 
&      \textcolor{lightgray}{.00} 
&       \textcolor{lightgray}{.00} 
&       \textcolor{lightgray}{.00} 
\\
$R_5$   &    &    &    &  .50 
&      \textcolor{red}{\textbf{.36}} 
&      \textcolor{orange}{.12} 
&      \textcolor{teal}{.01} 
&      \textcolor{lightgray}{.00} 
&      \textcolor{lightgray}{.00} 
&       \textcolor{lightgray}{.00} 
&       \textcolor{lightgray}{.00} 
\\
$R_6$   &        &        &        &        &      .50 
&      \textcolor{red}{\textbf{.20}} 
&      \textcolor{teal}{.02} 
&      \textcolor{lightgray}{.00} 
&      \textcolor{lightgray}{.00} 
&       \textcolor{lightgray}{.00} 
&       \textcolor{lightgray}{.00} 
\\
$R_7$   &        &        &        &        &        &      .50 
&      \textcolor{orange}{.10} 
&     \textcolor{teal}{.01} 
&      \textcolor{lightgray}{.00} 
&       \textcolor{lightgray}{.00} 
&       \textcolor{lightgray}{.00} 
\\
$R_8$   &        &        &        &        &        &        &      .50 
&       \textcolor{orange}{.12} 
&       \textcolor{orange}{.10} 
&       \textcolor{orange}{.10} 
&       \textcolor{teal}{.08} \\
$R_9$   &        &        &        &        &        &        &        &      .50 
&      \textcolor{red}{\textbf{.45}} 
&       \textcolor{red}{\textbf{.45}} 
&       \textcolor{red}{\textbf{.41}} \\
$R_{10}$   &        &        &        &        &        &        &        &        &      .50 
&       \textcolor{red}{\textbf{.50}} 
&       \textcolor{red}{\textbf{.45}} \\
$R_{11}$  &        &        &        &        &        &        &        &        &        &       .50 & \textcolor{red}{\textbf{.45}} \\
$R_{12}$  &        &        &        &        &        &        &        &        &        &         &       .50 \\
\end{tabular}
\end{footnotesize}
\ \\[2ex]
This example encourages to consider evaluations based on user feedback more carefully,
i.e. not to search for the only true ranking, but to weigh all possibilities against each other on the basis of their probabilities.

\section{Conclusion and Future Work}
Human uncertainty strongly influences the evaluation of recommender systems.
Hence, it is crucial to continue investigating this impact in our systems and evaluation processes.
In particular, this contribution is an opportunity to rethink about statistical soundness of even more modern and sophisticated quality measures than the RMSE. Future research may focus on the impact on other forms of recommender assessment and on developing new metrics that explicitly take human uncertainty into account.
 
\bibliographystyle{ACM-Reference-Format}
\bibliography{Literature} 

\end{document}